\documentclass[12pt,journal, draftclsnofoot,onecolumn]{IEEEtran}

\usepackage{savesym}
\usepackage{amsfonts,subfigure,multicol,color,verbatim,
graphicx,cite,epsfig,amssymb,amsmath,cases,bm,algorithm,
algorithmic,xcolor,multirow} 
\usepackage{amsmath}
\usepackage{array}
\usepackage{enumerate}
\usepackage{graphicx}
\usepackage{subfigure}

\savesymbol{iint}

\renewcommand{\IEEEQED}{\IEEEQEDopen}
\restoresymbol{TXF}{iint}

\IEEEoverridecommandlockouts

\begin{document}
\markboth{IEEE Networks} {Peng et al: Heterogeneous Cloud Radio
Access Networks\ldots}

\title{System Architecture and Key
Technologies for 5G Heterogeneous Cloud Radio Access Networks}

\author{
Mugen~Peng,~\IEEEmembership{Senior Member,~IEEE},
Yong~Li~\IEEEmembership{Member,~IEEE}, Zhongyuan~Zhao, and
Chonggang~Wang,~\IEEEmembership{Senior Member,~IEEE}

\thanks{Mugen~Peng (e-mail: {\tt pmg@bupt.edu.cn}), Yong~Li (e-mail: {\tt liyong@bupt.edu.cn}), and Zhongyuan~Zhao (e-mail: {\tt zyzhao@bupt.edu.cn}) are with the Key
Laboratory of Universal Wireless Communications for Ministry of
Education, Beijing University of Posts and Telecommunications,
China. Chonggang~Wang (e-mail: cgwang@ieee.org) is with the
InterDigital Communications, King of Prussia, PA, USA.}}

\renewcommand{\baselinestretch}{1.5}
\thispagestyle{empty} \maketitle \thispagestyle{empty}
\vspace{-15mm}
\begin{abstract}
Compared with the fourth generation (4G) cellular systems, the fifth
generation wireless communication systems (5G) are anticipated to
provide spectral and energy efficiency growth by a factor of at
least 10, and the area throughput growth by a factor of at least 25.
To achieve these goals, a heterogeneous cloud radio access network
(H-CRAN) is presented in this article as the advanced wireless
access network paradigm, where cloud computing is used to fulfill
the centralized large-scale cooperative processing for suppressing
co-channel interferences. The state-of-the-art research achievements
in aspects of system architecture and key technologies for H-CRANs
are surveyed. Particularly, Node C as a new communication entity is
defined to converge the existing ancestral base stations and act as
the base band unit (BBU) pool to manage all accessed remote radio
heads (RRHs), and the software-defined H-CRAN system architecture is
presented to be compatible with software-defined networks (SDN). The
principles, performance gains and open issues of key technologies
including adaptive large-scale cooperative spatial signal
processing, cooperative radio resource management, network function
virtualization, and self-organization are summarized. The major
challenges in terms of fronthaul constrained resource allocation
optimization and energy harvesting that may affect the promotion of
H-CRANs are discussed as well.
\end{abstract}


\newpage


\section{Introduction}
With the rapid development of mobile internet and internet of things
(IoTs), the demands for high-speed data applications, such as
high-quality wireless video streaming, social networking and
machine-to-machine communication, have been growing exponentially
recently. It is envisioned that the total daily mobile traffic in
the representative Western European countries will grow 67 times
from 186 terabyte (TB) to 12540 TB through 2010 to 2020, and the
total worldwide mobile traffic of 351 exabyte (EB) in 2025
represents a 174\% increase compared with
2020\textcolor[rgb]{1.00,0.00,0.00}{\cite{UMTS}}. Currently, the
cellular networks including the first generation (1G), second
generation (2G), third generation (3G) and fourth generation (4G)
are far from satisfying the significant traffic increments and the
high energy efficiency (EE) because a lot of power of a base station
(BS) is used to overcome path loss, which in turn causes
interferences to other users. The fifth generation (5G) system
deployed initially in 2020 is expected to provide about 1000 times
higher wireless area capacity and save up to 90\% of energy
consumption per service compared with the current 4G system. More
than 1000 Gbit/s/km$^2$ area spectral capacity in dense urban
environments, 10 times higher battery life time of connected
devices, and 5 times reduced end-to-end (E2E) latency are
anticipated in 5G systems. The new 5G air interface and spectrum
should be combined together with the long term evolution (LTE) and
WiFi to provide universal high-rate coverage and a seamless user
experience\textcolor[rgb]{1.00,0.00,0.00}{\cite{5G}}.

To achieve these goals of 5G systems, advanced radio access
technologies and all-internet protocol (IP) open internet network
architectures should be evolved smoothly from 4G
systems\textcolor[rgb]{1.00,0.00,0.00}{\cite{ITU2020}}. Accurately,
the new breakthroughs in the baseband and radio frequency (RF) are
required to enable computationally intensive and adapt to new air
interfaces in 5G systems. A significant and advanced baseband
computation is required to meet the complex requirements of new
solutions like large-scale cooperative signal processing in the
physical layer. Meanwhile, the new breakthroughs in the integrated
access node and heterogeneous convergence are required to enable the
ultra dense radio nodes to work efficiently. The plug-and-play
function becomes essential to commercial deployments, in which the
available spectral resources should be allocated and the
corresponding parameters should be self-organized. Furthermore, the
software-defined air interface technologies should be seamlessly
integrated into the 5G radio access network (RAN) architectures. The
cloud computing based radio access infrastructures would provide
on-demand resource processing, delay-aware storage, and high network
capacity wherever needed.

\subsection{5G C-RAN Solution}

Some advanced technologies, such as the cloud radio access network
(C-RAN) and ultra small cells based heterogeneous network (HetNet),
have been presented as potential 5G solutions. C-RAN has attracted
intense research interests from both academia and industry (such as
China Mobile, Huawei, Alcatel Lucent,
Qualcomm)\textcolor[rgb]{1.00,0.00,0.00}{\cite{bib:CRAN}}. In
C-RANs, a large number of low-cost remote radio heads (RRHs) are
randomly deployed and connect to the base band unit (BBU) pool
through the fronthaul links. C-RANs have several advantages: First,
by moving RRHs closer to the users, a higher system capacity and
lower power consumption can be achieved because the signal doesn't
need to propagate a long distance to reach the users; Second, since
the baseband processing is centralized at the BBU pool, the
cooperative processing techniques to mitigate interferences can be
leveraged; Third, by exploiting the resource pooling and statistical
multiplexing gain, C-RAN is much more efficient in both energy and
cost aspects because it is needless to dimension the computing
resource of each traditional BS according to the individual peak
load. However, the fronthaul constraints have great impact on
worsening performances of C-RAN, and the scale size of RRHs
accessing the same BBU pool is limited and could not be too large
due to the implementation complexity. Furthermore, many kinds of
system architectures have been proposed by different mobile
operators, manufactories and even researching institutes to explore
potential advantages of C-RANs. Therefore, an unified C-RAN for 5G
is still not straightforward.

\subsection{5G HetNet Solution}

To increase the capacity of cellular networks in dense areas with
high traffic demands, the low power node (LPN) serving for the pure
``data-only'' service with high capacity is identified as one of key
components in
HetNets\textcolor[rgb]{1.00,0.00,0.00}{\cite{bib:HetNet}}. One key
advantage of HetNets is to decouple the control plane and user
plane. LPNs only have the control plane, while the control channel
overhead and cell-specific reference signals of LPNs can be fully
shifted to macro base stations (MBSs). Unfortunately, an underlaid
structure that MBSs and LPNs reuse the same spectral resources could
lead to severe inter-tier interferences. Hence, it is critical to
suppress interferences through advanced signal processing techniques
to fully unleash the potential gains of HetNets, such as adopting
the advanced coordinated multi-point (CoMP) transmission and
reception technique to suppress both intra-tier and inter-tier
interferences. It was reported that the average spectral efficiency
(SE) performance gain from the uplink CoMP in downtown Dresden field
trials was only about 20 percents with non-ideal backhaul
in\textcolor[rgb]{1.00,0.00,0.00}{\cite{bib:CoMPgain}}.

\subsection{5G H-CRAN Solution}

To fulfill new breakthroughs anticipated in 5G systems, and overcome
the aforementioned challenges in both C-RANs and HetNets, we
presented heterogeneous cloud radio access networks (H-CRANs) as 5G
RANs in our prior
work\textcolor[rgb]{1.00,0.00,0.00}{\cite{bib:peng_HCRAN}}, which
are fully backward compatible with different kinds of C-RANs and
HetNets. The motivation of H-CRANs is to embed the cloud computing
technology into HetNets to realize the large-scale cooperative
signal processing and networking functionalities, and thus SE and EE
performances are substantially improved beyond existing HetNets and
C-RANs. In the H-CRAN based 5G system, the control plane and user
plane are decoupled. MBSs are mainly used to deliver control
signalling of the whole H-CRAN and provide the seamless coverage,
while RRHs are used to provide the high speed data transmission in
the hot spots with huge data services. The large-scale cooperative
signal processing and networking techniques in H-CRANs are
threefold. The first comprises advanced spatial signal processing
techniques in the physical layer (PHY), including the centralized
massive multiple-input-multiple-output (MIMO) and the distributed
large-scale spatial cooperative processing; the second comprises
large-scale cooperative radio resource management and cloudization
in the medium access control (MAC) and upper layers; the third
comprises intelligent and self-organizing network management in the
network layer to support self-configuration, self-optimization, and
self-healing in the ultra dense communication scenario with huge
number of nodes deployed randomly.

\subsection{Motivation and Organization}

In this article, we are motivated to make an effort to offer a
comprehensive discussion on recent advances of system architectures
and technological principles in 5G H-CRANs. Specifically, the
application scenarios and H-CRAN system architecture are presented,
where the new communication entity \textbf{Node C} is defined and
the software-defined H-CRAN architecture is designed. The
large-scale cooperative signal processing and networking techniques
strictly related to improve SE and EE in H-CRANs are surveyed,
including advanced spatial signal processing, cooperative radio
resource management, network function virtualization (NFV), and
self-organizing network (SON). The challenging issues related to
resource allocation optimization and energy harvesting are discussed
as well.

The remainder of this paper is outlined as follows. The application
architecture and system components of H-CRANs will be introduced in
Section II. The promising large-scale cooperative signal processing
and networking technologies will be shown in Section III. Future
challenges will be highlighted in Section IV, followed by the
conclusions in Section V.

\section{Application Architecture and System Components}

\vspace*{15pt}

The H-CRAN based 5G systems have capabilities to provide anytime,
anywhere gigabit data rate service to the desired user equipments
(UEs), where an ``edgeless" experience to UEs is provided under the
help of heterogeneous MBSs and ultra dense RRHs. Based on cloud
computing technologies, the on-demand resource processing, storage
and network capacity wherever needed are fulfilled. Software-defined
air interfaces and networking technologies are seamlessly integrated
into the H-CRAN architectures, which enables the flexibility to
create new services and applications.

\subsection{Application Architecture of H-CRANs}

\begin{figure}
\centering
\subfigure[Application scenarios of 5G systems]{ \label{fig2} 
\includegraphics[width=3.3in]{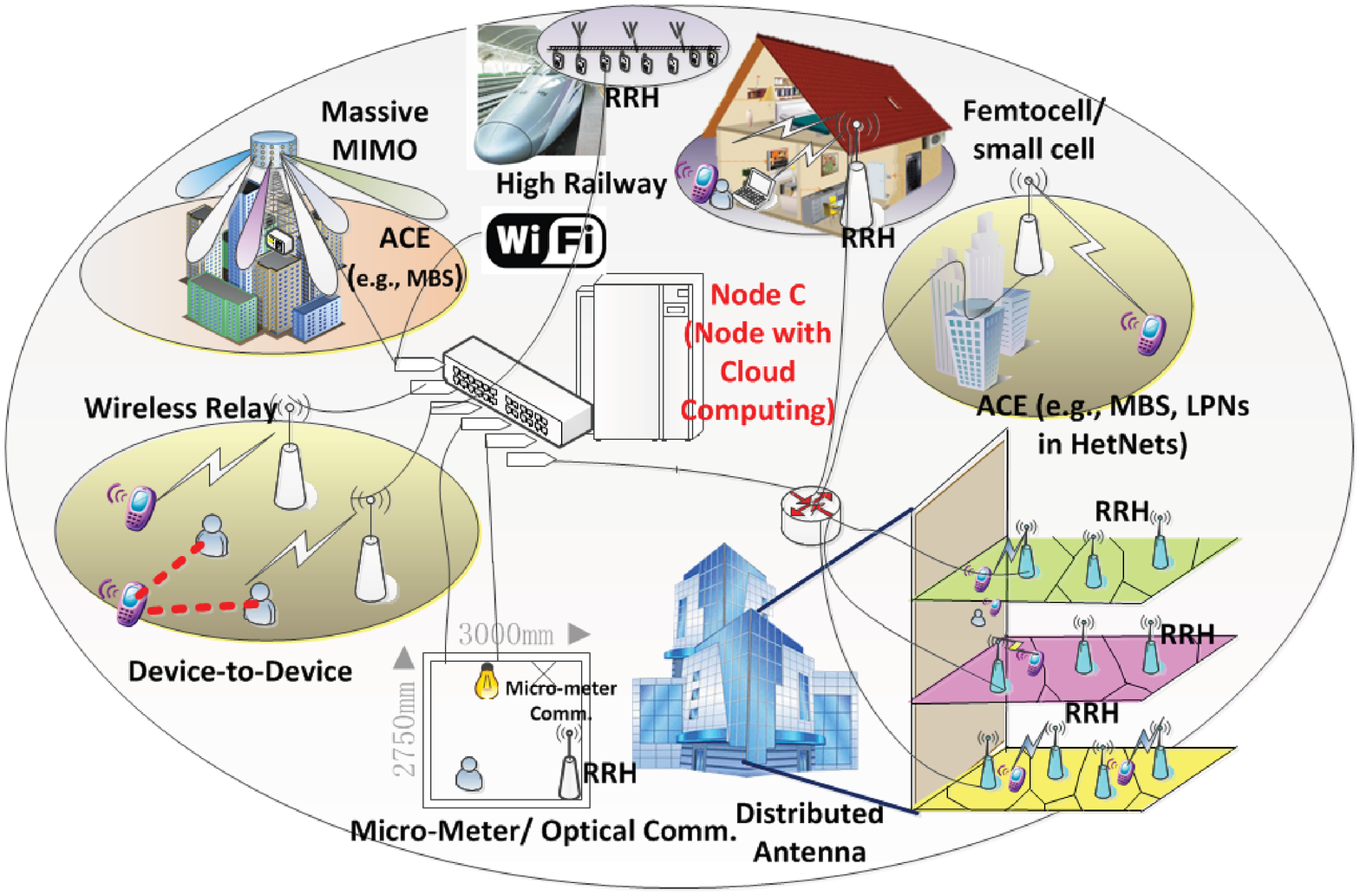}}
\hspace{0.01in}
\subfigure[5G components
and key functionalities]{ \label{fig3} 
\includegraphics[width=3.3in]{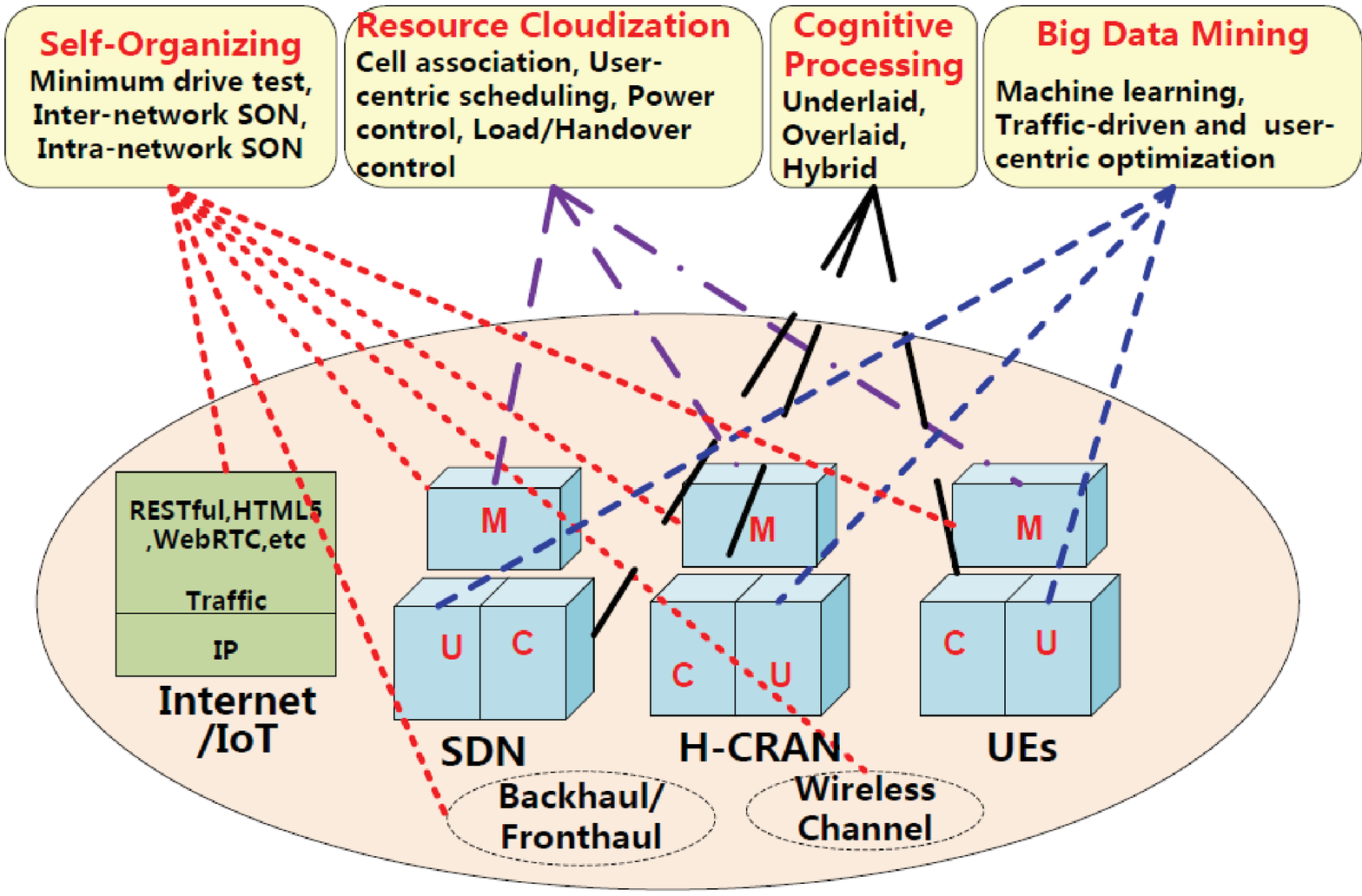}}
\caption{\textbf{Application architecture, system components, and
key functionalities in H-CRANs}}
\label{fig1} 
\end{figure}

H-CRANs shown in Fig. \ref{fig2} take full advantages of the cloud
computing and heterogeneous convergence technologies, where the new
communication entity \textbf{Node C} (Node with cloud computing) as
the evolution of BS in the $3^{rd}$ generation project partnership
(3GPP) is presented to converge different RANs for the existing
ancestral communication entities (ACEs, i.e., MBSs, micro BSs, pico
BSs, etc.) and offer processing and networking functionalities in
physical and upper layers for the new designed RRHs. When Node C
works to converge ACEs, it can be regarded as a convergence gateway
to execute the cooperative multiple-radio resource managements
(CM-RRM) and media independent handover (MIH) functionalities, and
the functionalities of traditional radio network controller (RNC)
and BS controller (BSC) can be embodied into Node C. When Node C is
used to manage RRHs, it acts as the BBU pool, which is inherited
from C-RANs. In addition, Node C has a powerful computing
capabilities to execute the large-scale cooperative signal
processing in the physical layer and large-scale cooperative
networking in the upper layers.

In H-CRANs, RRHs are mainly used to provide the high speed data
transmission without the control plane in hot spots. The control
channel overhead and cell specific reference signals for the whole
H-CRAN are delivered by ACEs. To guarantee the seamless coverage,
UEs nearer to ACEs than RRHs are served by ACEs and called HUEs. To
improve the intelligences and self-organizing capabilities, the
universal plug and play functionality is critical, which allows both
the new deployed RRHs and the existing ACEs to access the Node C
instantaneously and automatically. With the help of Node C in
H-CRANs, hundreds of RRHs and several tens of ACEs can be served
simultaneously, and the control and user plane separation can be
realized efficiently.

In view of the problems existing in internet protocol (IP) networks,
software-defined network (SDN) is proposed as the IP evolution in 5G
systems to simplify the network construction and operational costs
by adopting the NFV
technology\textcolor[rgb]{1.00,0.00,0.00}{\cite{bib:SDN}}. The SDN
based wireless network infrastructures are constantly emerging, in
which SDN is co-located with the internet/IoT network entities and
decentralizes RRHs/ACEs closer to the desired UEs. To be compatible
with the SDN architecture, the adaptive signaling/control mechanism
between connection-oriented and connectionless is supported in
H-CRANs, which can achieve significant overhead savings in the radio
connection/release by moving away from a pure connection-oriented
mechanism. Node C is interfaced to SDN to complete NFV
functionalities, which is urgent to be adaptive to the dramatically
growing traffic features and demands. Different kinds of radio
communication technologies in the physical layer can be adopted in
RRHs to improve transmission bit rates, such as IEEE 802.11 ac/ad,
millimeter wave communication, and even optical light communication.

To further decrease the power consumption in H-CRANs, the activating
RRHs are adaptive to the traffic volume. When the traffic load is
low, some potential RRHs fall into sleep under supervising of Node
C. However, when the traffic load becomes tremendous in a small
special zone, both the heterogeneous ACEs and dense RRHs work
together to satisfy the huge capacity requirements. Further, based
on the NFV technology, RRHs with high load can borrow the radio
resources of neighboring homogeneous and heterogeneous RRHs with low
load.

\subsection{System Components of H-CRANs}

As shown in Fig. \ref{fig3}, the H-CRAN based 5G system is comprised
of UEs, H-CRAN, SDN and internet/IoTs. There exists three planes in
this architecture: user (U) plane, control (C) plane and management
(M) plane. Each of them carries out different areas of operations
and processes different types of functionalities. Generally
speaking, the U plane carries the actual user traffic and executes
the related traffic processing to satisfy various quality of service
(QoS) requirements. The C plane carries the control signaling, and
takes charge of the resource allocation and traffic processing to
improve SE and EE. The M plane executes administration and
operation, and is mainly responsible for the adding, deleting,
updating, and modification of the logic and interactions for U and C
planes. Except for internet/IoT networks, the other three
components, such as UE, H-CRAN and SDN, can be vertically divided
into U, C and M planes. It is worth noting that only C plane is
configured in RRHs, while both C and U planes are incorporated in
ACEs.


The H-CRAN architecture is software defined, and is characterized by
attributes of both SDN and cloud computing. The network control
information is delivered from SDN to Node C through the standardized
southbound interface, i.e,
OpenFlow\textcolor[rgb]{1.00,0.00,0.00}{\cite{bib:SDN}}. By
centralizing the network intelligence, the decision-making is
facilitated from a global viewpoint. These features are opposed to
the current wireless networks, which are built on an autonomous
view, and nodes are unaware of the overall state of the network
wherein. Meanwhile, the presented architecture is inherently
controlled by the software defined functionality. The centralized
programmable architecture enables the network configuration to be
automated under the rapid adoption of the cloud computing. By
providing open application programming interfaces (APIs) for
internet/IoT applications to interact with the other entities in 5G
systems, H-CRAN can achieve unprecedented innovation and
differentiation. This software-defined H-CRAN architecture brings
significant benefits compared with traditional architectures. On the
one hand, the programmability of logically centralized controller
makes it convenient to incorporate new ideas in the network and
offers network managers the flexibility to configure, manage, and
optimize network resources via dynamic, self-organizing programs. On
the other hand, benefitting from the centralized approach, operators
can make network-wide traffic forwarding decisions in the logically
centralized controller instead of configuring all network devices
individually to make changes in network behavior.

To improve SE and EE performances of the software-defined H-CRAN,
there are 4 key functionalities to be implemented, such as the
self-organizing, radio resource cloudization, cognitive processing,
and big data mining. The big data mining functionality imports the
machine learning technology into the cooperative signal processing
and intelligent networking, which makes the 5G system adaptive to
the Internet/IoT traffic's features and demands. The H-CRAN based 5G
systems are traffic-driven and user-centrical. To avoid the
inter-tier interference, the cognitive processing technique is
utilized to make RRHs and ACEs work cooperatively in the overlaid
scenario when the overall system load is not high. Furthermore, to
improve the SE performance when the overall system load is high, the
radio resource cloudization technique is used to decrease the
inter-tier interference and improve the flexibility of radio
resource's reuse. To decrease the operation cost and enhance the
networking intelligences, the self-organizing H-CRAN is necessary.
With measurement and analysis of SE and EE performances ,
self-organizing functionality automatically configures and optimizes
the traffic, fronthaul and radio resources, and regulates operations
of the M plane without human interventions.

\section{Promising Key Technologies}

To take full advantages of H-CRANs, the large-scale cooperative
spatial signal processing (LS-CSSP) and cooperative networking
techniques should be exploited. The LS-CSSP technique is utilized to
fulfill the interference cancelations through collaboration over the
software-defined architectures. The user-centric and delay-aware
large-scale cooperative radio resource management (LS-CRRM) should
be highlighted. Further, to be compatible with the SDN based 5G core
network, the heterogeneous radio resources should be reused and
cloudized, and NFV should be exploited and realized. Considering the
complex and costly network planning and maintaining, the large-scale
self-organizing network (LS-SON) is indispensable to improve
intelligences and lower costs. Some other technologies impacting on
performances of H-CRANs are not introduced in this Section, such as
the fronthaul compression and quantization, dynamical cell
clustering, and channel estimation, are still challenging topic for
research in the future.

\subsection{Large-Scale Cooperative Spatial Signal Processing (LS-CSSP)}

\begin{figure}[!htp]
\centering
\includegraphics[width=6.0in]{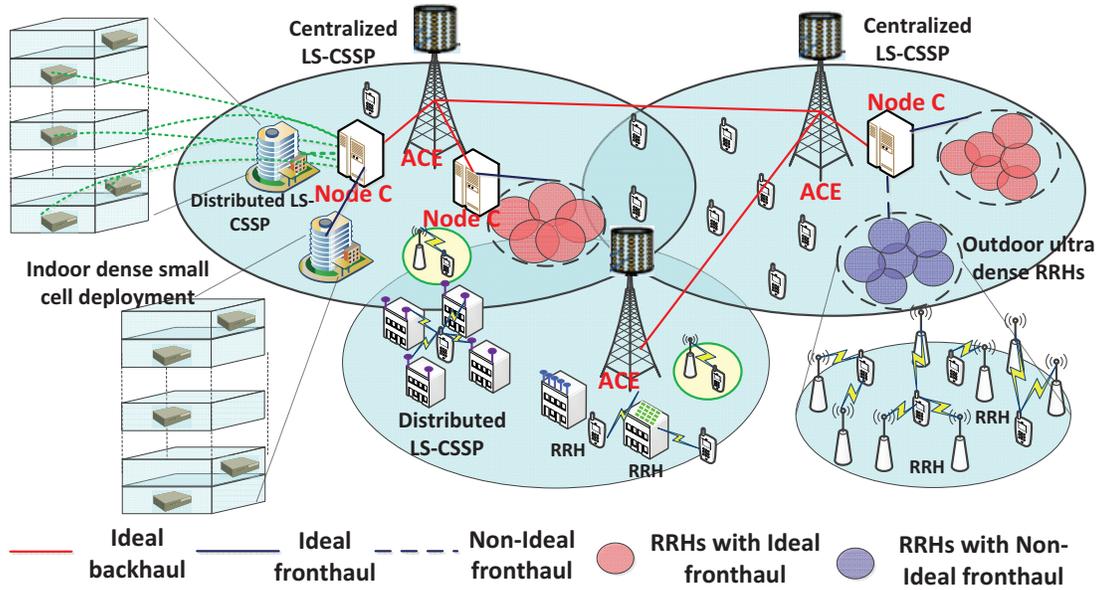}
\caption{\textbf{Centralized and distribution LS-CSSP in
H-CRANs}}\label{LS-CSSP}\vspace*{-1em}
\end{figure}

For the ultra dense nodes deployed in the traditional cellular
networks or HetNets, although the impact of large-scale fading is
alleviated, the existence of several co-channel interferences
becomes the performance bottleneck. The LS-CSSP, including both
distributed and centralized modes, is prominent to improve SE
performance with the gigabit rate as shown in Fig. \ref{LS-CSSP}.
The distributed LS-CSSP relies on the spatial multiplexing, which in
turn shares the same radio resources amongst RRHs. By transmitting
collaboratively among multiple RRHs, which can be treated as a
virtual antenna array, the interference can be avoided spatially
without occupying extra radio resources.

\begin{figure}[!htp]
\centering
\includegraphics[width=4.0in]{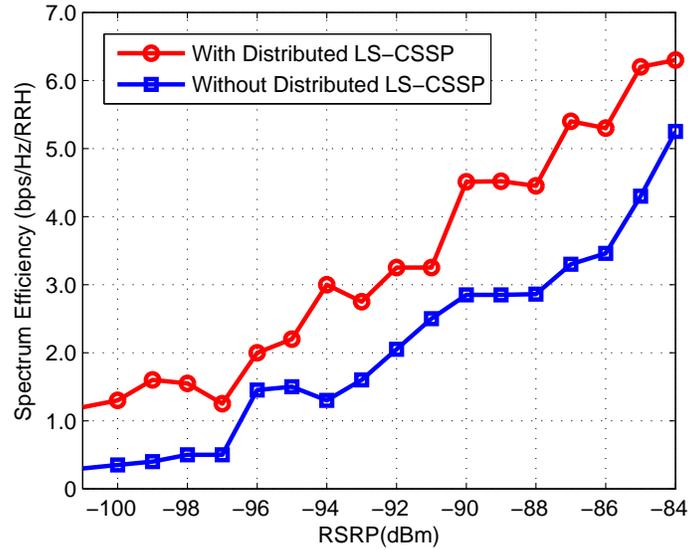}
\caption{\textbf{Uplink Throughput Gains from distributed LS-CSSP,}
where the inter-site distance is around 1 kilometer, 2 cells are
dynamically chosen to perform the uplink LS-CSSP in the
collaborative cluster with 6 cells, and 5 interfering users in the
surrounding cells with 50 \% resource block (RB) utilization ratio
are
configured\ref{LS-CSSP_performance}\textcolor[rgb]{1.00,0.00,0.00}{\cite{UplinkCOMP}}.}\label{LS-CSSP_performance}\vspace*{-1em}
\end{figure}

For the distributed LS-CSSP, it is necessary to exploit cooperative
precoding schemes that make special considerations of capacity
constraints on fronthaul, implementation complexity, and difficulty
in obtaining perfect channel state information (CSI) for all
connected RRHs and ACEs. Though the dirty paper pre-coding can
achieve a good capacity region, it is impractical because the
complication increases with the number of connected RRHs and ACEs.
The linear precoding is a good alternative with less complexity,
which can be optimized jointly with combination of constraints on
fronthaul\textcolor[rgb]{1.00,0.00,0.00}{\cite{bib:precoding}}.
Meanwhile, the cell association has great impact on precoding
performances, the joint cell selection and precoding is key to
improve EE and SE performances. As shown in Fig.
\ref{LS-CSSP_performance}\textcolor[rgb]{1.00,0.00,0.00}{\cite{UplinkCOMP}},
in the uplink, the distributed LS-CSSP has good potential to
strengthen uplink signal and suppress inter-RRH interference, whose
gain can reach $40 \sim 100\%$ in the weak coverage area (that is
the area where reference signal receive power (RSRP) is lower than
-95dBm), and in some area the gain can exceed $100\%$. Due to the
severe inter-RRH interference, the uplink throughput cannot reach
peak at good coverage area (e.g. where RSRP higher than -90dBm) when
the distributed LS-CSSP is not utilized. Distributed LS-CSSP can
bring $1 \sim 2$ Mbps throughput boost and result in $20 \sim 50\%$
throughput gain at the test area.

The superior centralized processing capability of H-CRAN provides
great convenience for making distributed LS-CSSP effective, even can
support the combination with other advanced techniques, such as
interference alignment. However, some existing technical issues are
still challenging. One of key challenging problems is to determine
the scale size of distributed LS-CSSP adaptively. Due to the
requirement of a large amount of CSIs, there exists a tradeoff
between the SE gain and scale size of the distributed LS-CSSP. On
the one hand, the improvement of SE performance might be limited
when the scale of the distributed LS-CSSP is small. On the other
hand, by allowing the participants overly, the signaling exchanging
is intractable, and the accuracy and instantaneity of CSI decline as
well, which degrades SE performances severely. Therefore, it is
critical to clarify the boundary conditions in which the distributed
LS-CSSP can achieve significant gains in H-CRANs.

By enlarging the special degrees of freedom, MIMO technique can
improve SE and EE performances via achieving an extra
diversity-multiplexing tradeoff. The centralized LS-CSSP is equipped
with hundreds of low-power antennas at a co-located ACE site, which
is mainly used to improve capacity, extend coverage, and decrease
antenna deployment complexity. According to the law of large
numbers, the channel propagation condition can be hardened, which
ensures that the transmission capacity increases linearly as the
number of antennas increases. The centralized LS-CSSP can increase
the capacity 10 times or more, and improve the radiated EE
performance on the order of 100 times as
well\textcolor[rgb]{1.00,0.00,0.00}{\cite{bib:LS-MA}}.

Unlike RRHs, ACEs do not need to upload all their observations to
Node C for the joint signal cooperative processing. By deploying
ACEs with the centralized LS-CSSP, instead of using a huge number of
RRHs in some coverage areas, the overload on fronthauls between RRHs
and Node C is released, and the time latency of backhaul between
MBSs and Node C is alleviated. Compared with ACEs without the
centralized LS-CSSP, or the conventional C-RAN configuration, ACEs
with the centralized LS-CSSP reduce inter-tier interferences to
adjacent RRHs/ACEs because they can serve a large area with diluting
the density of active RRHs. Moreover, the cooperative beamforming
between ACEs and RRHs can suppress remaining inter-tier
interferences.

\subsection{Large-Scale Cooperative Radio Resource Managements (LS-CRRM)}

\begin{figure}
\centering
\subfigure[Enhanced soft
fractional frequency reuse]{ \label{fig5_a} 
\includegraphics[width=2.7in]{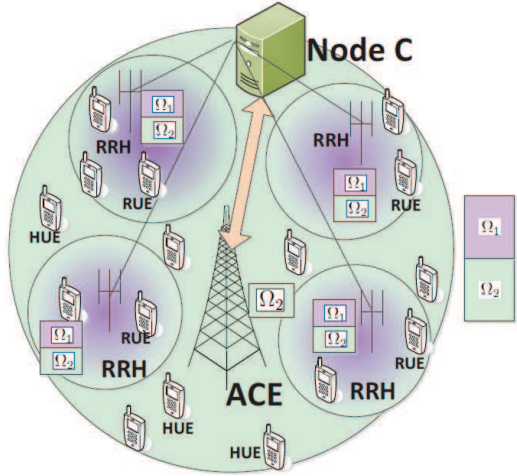}}
\hspace{0.01in}
\subfigure[Average queue length vs mean arrival rate of RUEs]{ \label{fig5_b} 
\includegraphics[width=3.7in]{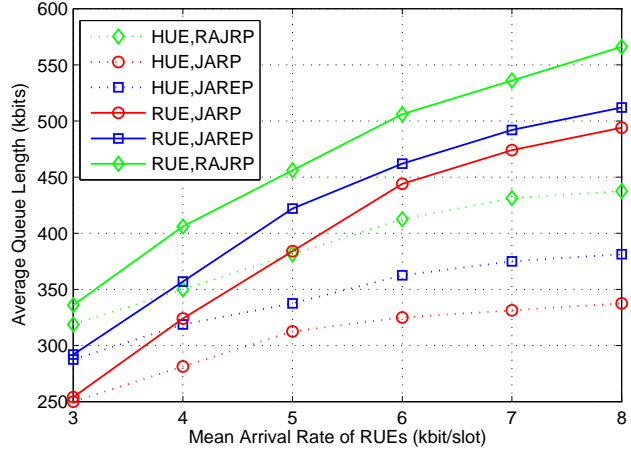}}
\caption{\textbf{Enhanced soft FFR and performance evaluations of
LS-CRRM schemes in H-CRANs}, where H-CRAN consists of 1 ACE and 4
RRHs, the number of RUEs is 4, the number of HUEs is 8, the number
of RBs is 10, the bandwidth of each RB is 15kHz, and the mean
traffic arrival rate of HUEs as 3
kbits/slot\textcolor[rgb]{1.00,0.00,0.00}{\cite{LIjian_CRRM}}.}
\label{fig1} 
\end{figure}

To optimize performances of H-CRANs, more advanced radio resource
management schemes are required than the traditional cellular
network due to the additional computing resources. Since the
physical radio resources can be completely reused among RRHs, the
inter-tier interference between RRHs and ACEs should be suppressed
with advanced LS-CRRM techniques. The traditional soft fractional
frequency reuse (S-FFR) is considered to be an efficient inter-cell
and inter-tier interference coordination technique, in which the
service area is partitioned into spatial regions, and each subregion
is assigned with different frequency sub-bands. Therefore, the
cell-edge-zone UEs do not interfere with cell-center-zone UEs, and
with an efficient channel allocation method, the cell-edge-zone UEs
may not interfere with neighboring cell-edge-zone
UEs\textcolor[rgb]{1.00,0.00,0.00}{\cite{SFR}}. Considering that the
ACE is mainly used to deliver the control signalling and guarantee
the seamless coverage, the QoS requirement of HUEs is not high,
while the QoS requirement for UEs accessing RRHs (denoted as RUEs)
is with a high priority. Consequently, an enhanced S-FFR scheme
should be proposed to suppress the inter-tier interference between
the ACE and RRHs in Fig. \ref{fig5_a}, where only partial radio
resources are allocated to both RUEs and HUEs with the low QoS
requirement, and other radio resources are allocated to RUEs with
high QoS requirements.

For the enhanced S-FFR, the RUEs with low QoS requirements will
share the same radio resources with HUEs, which is absolutely
different from that in the traditional S-FFR, where the radio
resources allocated to cell-edge-zone UEs are orthogonal. If the
traditional S-FFR is utilized in H-CRANs, only the cell-center-zone
RUEs will share the same radio resources with HUEs, which decreases
the SE performance significantly. Further, how to determine UEs
located in the cell-edge or cell-center zone is challenging for the
traditional S-FFR. Fortunately, these problems are solved via the
enhanced S-FFR, where only the QoS requirement should be
distinguished for RUEs. To avoid inter-tier interference between
RRHs and ACEs, the outband frequency is preferred to be used
according to the standards of HetNets in 3GPP, which means that RBs
for ACEs should be different from those for RRHs. Meanwhile, to save
the occupied frequency bands, the inband strategy is also defined in
3GPP, which means that both RUEs and HUEs will share the same RBs
even when the inter-tier interference exists. To be completely
compatible with inband and outband strategies in 3GPP, only two RB
sets (namely ${\Omega _1}$ and ${\Omega _2}$) should be divided for
the enhanced S-FFR scheme. Obviously, if the locations of RUEs could
be known and the traffic volume in different zones are clearly
anticipated, more RB sets can be divided in ${\Omega _1}$ to achieve
a higher performance gain. ${\Omega _1}$ is only allocated to RUEs
with high rate-constrained QoS requirements, and ${\Omega _2}$ is
allocated to RUEs and HUEs with low rate-constrained QoS
requirements. Since all signal processing for different RRHs are
executed on the BBU pool centrally, the inter-RRH interferences can
be avoided and the same radio resource can be shared amongst RRHs,
thus the radio resources among RRHs are reused and virtualized.
Based on the enhanced S-FFR, the inter-tier interference among RRHs
and ACEs can be suppressed efficiently.

To avoid the traffic congestion and obtain the near-optimal
performances in H-CRANs, the LS-CRRM scheme should take the queue
state information (QSI) into account to guarantee the fairness among
UEs and improve the utilization efficiency of radio resource.
Particularly, the joint optimization problem taking traffic
admission control, RRH/MBS association, RB and power allocation into
account subject to the average and instantaneous power consumption
constraints should be researched, and performances of the
corresponding joint-association-RB-power allocation (JARP) scheme
are evaluated in Fig. \ref{fig5_b}, where the
random-association-joint-RB-and-power allocation (RAJRP) scheme and
joint-association-and-RB-equal-power allocation (JAREP) scheme are
compared as
baselines\textcolor[rgb]{1.00,0.00,0.00}{\cite{LIjian_CRRM}}. For
RAJRP, the HUE randomly accesses the target RRH with the probability
proportional to the throughput virtually, and the RB and power are
jointly optimized using our algorithm. For JAREP, association and
RBs are jointly optimized using our algorithm and the maximum
average power is equally allocated on all RBs of each RRH.

The performances of average queue length of HUEs and RUEs versus
mean traffic arrival rate of RUEs are compared among these three
schemes. It is observed that the bigger the traffic arrival rate of
RUEs, the larger the average queue length for HUEs. This can be
understood by the fact that more RBs and power are allocated to RUEs
to maintain the queue stability, leaving less resource for HUEs
which are intended to access RRHs. As the mean arrival rate of RUEs
increases, the system throughput utility increases, but with a
diminishing slope. The rationale is that all traffic arrivals of
RUEs can be admitted by these schemes according to the
threshold-based admission control policy when the traffic arrival
rate of RUEs is relatively small, and some traffic arrivals for RUEs
have to be denied to stabilize the traffic queues with higher
traffic arrival rate of RUEs. JARP achieves a higher utility and
smaller average queue length than JAREP and RAJRP under all traffic
arrival rates, which implies the importance of the joint
optimization of association, RB, and power allocation.

\subsection{Network function virtualization (NFV) in H-CRANs}

Based on the software-defined H-CRAN architecture, NFV is an
emerging technology for cost-effective sharing of scarce radio and
computing resources in 5G systems, which aims to evolve standard
computing virtualisation technology to consolidate network
equipments into high volume servers, switches and storage.
Accurately, as highly complementary to the control and user plane
separation in H-CRANs, NFV in SDNs is applicable to the packet
processing in the U plane and the control function in the C plane,
which offers the potential capabilities for both enhancing service
delivery and reducing overall costs. By enabling NFV with
OpenFlow-enabled SDN, network operators can realize significant
benefits through utilizing the cloud computing technology.

\begin{figure}[!htp]
\centering
\includegraphics[width=4.5in]{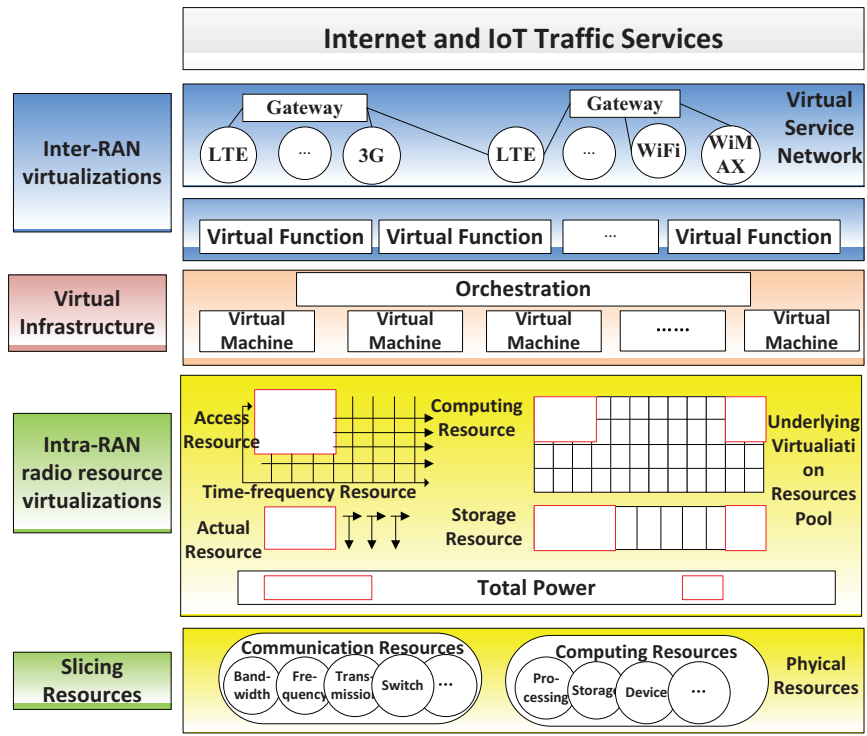}
\caption{\textbf{Network function virtualization (NFV) in H-CRANs}
}\label{Fig7_RV}\vspace*{-1em}
\end{figure}

The H-CRAN infrastructures provide potential capabilities to enhance
resource availability and usage by means of orchestration and
management mechanisms, which are applicable to the automatic
instantiation of virtual appliances and the management of resources
by assigning virtual appliances to the correct computing core,
memory and interfaces. The NFV designed for H-CRANs is shown in Fig.
\ref{Fig7_RV}, where the physical radio resources are cloudized, and
computing resources in both intra-RANs and inter-RANs are
virtualized. The functionalities of the virtual machine (VM)
technology executed in Node C include: (1) Allow for quick creation
and deletion of VM instances for management purposes; (2) Support a
wide variety of distributions to allow customization of environment;
(3) Allow the user to customize everything from the Kernel to the
drivers; and (4) Provide means for easy administration. To fulfill
the virtual infrastructure, the Node C should embody the generic
high-volume computing servers, storage devices, high-volume network
switches and be organized by the orchestration.

An important aspect for the NFV is the transparency, i.e., the
virtual nodes cannot see or exchange any type of information, in
order to assure isolation of the heterogeneous convergence. By using
the virtual function in Fig. \ref{Fig7_RV}, the heterogeneous of
ACEs and RRHs are virtualized and are transparent to the lower
layers. Additionally, the data exchange in the virtual layer is
transparent to the physical provider to preserve the privacy of
customers. Nevertheless, some minimal primitives to inspect the
activity of different slices are normally available. As an example,
the controller of physical radio resources is allowed to know the
actual usage of computational resources and traffic consumption.

The design and implementation of NFV in 5G systems should highlight
the coexistence of slices with bandwidth-based and resource-based
reservations because the bandwidth achieved by a slice from a given
amount of resources varies with the channel quality of the users and
the flow scheduling policies. H-CRANs have to consider resource
sharing for uplink traffic, while UEs originate the uplink traffic
randomly, which makes it hard to achieve the two conflicting goals
of isolation and efficient resource utilization across slices.
Meanwhile, H-CRANs often incur considerable overheads due to
signaling and retransmissions that have to be properly accounted
for, which hurts the resource utilization and flexibility.

\subsection{Large-Scale Self-Organizing H-CRANs (LS-SON)}

To fulfill the universal plug and play functionality, offload the
traffic from the core network and manage computing and radio
resources more efficiently using the service-aware controls, the
large-scale self-organizing functionnaires are critical to guarantee
the giant RRHs and ACEs working in the intelligent manner. SON was
proposed to reduce operational costs for service providers in LTE
cellular systems and
HetNets\textcolor[rgb]{1.00,0.00,0.00}{\cite{bib:SON2}}. Considering
that too many parameters should be configured and optimized due to
the combinations of heterogeneous convergence and cloud computing in
H-CRANs, and that the radio resources are shared and virtualized,
LS-SON is the key to integrate ultra learning, ultra planning, ultra
configuration, and ultra optimization into a unified automatic
process requiring minimal manual interventions with the
centralization of cloud computing. The LS-SON not only reduces the
complexity of managing co-channel interferences in H-CRANs, but also
saves operational costs to all RRHs and ACEs. LS-SON is used to
harmonize the whole network management approaches and improve the
overall operational efficiency. On the other hand, the availability
of LS-SON solutions lead to identify powerful optimization
strategies, and suppress co-channel interferences and improve both
EE and SE performances.

\begin{figure}[!htp]
\centering
\includegraphics[width=5.0in]{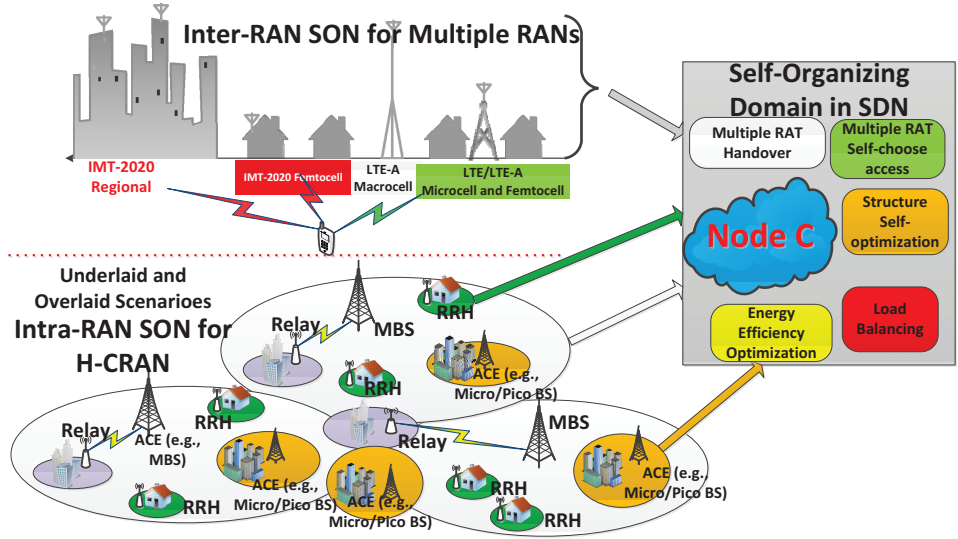}
\caption{\textbf{Self-organizing H-CRANs}
}\label{Fig8_SON}\vspace*{-1em}
\end{figure}

Due to the existence of centralized Node C co-located with the SDN
in the H-CRAN based 5G systems, the self-configuration,
self-optimization, and self-healing functionalities are implemented
in the centralized SON architecture. Since Node C needs to converge
multiple RANs and process the RRHs with cloud computing
cooperatively, which is shown in Fig. \ref{Fig8_SON}, the inter-RAN
and intra-RAN SON functionalities should be implemented in Node C.

For the self-configuration, since RRHs are utilized to support
transmissions with high data rates, the physical cell identifier
(PCI) assignment is not necessary for RRHs, and only the radio
resource should be self-assigned to each RRHs intelligently.
However, the self-configuration of both PCI and radio resource
should be fulfilled in ACEs. Thanks to Node C, these
self-configuration cases for ACEs are handled in the centralized
mode. For the self-optimization cases, energy saving and mobility
load balance (MLB), mobility robustness optimization (MRO) are three
key cases. To save the energy consumption, the dynamic energy saving
strategy is used to optimize the number of hardware units in every
moment, so that their number is minimized while the desired QoS is
still ensured. In other words, switching off RRHs and ACEs when they
are not needed. The MLB has been envisaged as a distributed
functionality, in which the algorithms are executed in the BS in 4G
systems. However, the MLB will be located in Node C in a centralized
mode. For the self-healing, the perception that a problem is
occurring or about to occur is the first step, followed by finding
the root cause of the detected problem, and the final step is to
identify and apply the appropriate corrective actions to (fully or
partially, definitely or temporarily) restore the service. All these
aforementioned LS-SON functions should be designed and fulfilled in
the future.

\section{Challenges and Open Issues in H-CRANs}

Although there have been some progresses and initial achievements in
the above mentioned potential key techniques for H-CRANs, there are
still many challenges ahead, including the optimal resource
allocation over the constrained fronthaul, energy harvesting,
backward compatible standard development, etc. Some classical
challenges are discussed as samples in this section.

\subsection{Fronthaul Constraints and Performance Optimization}

The wire and wireless fronthauls are becoming increasingly an
important infrastructure in H-CRANs, which not only affects the data
throughput available to UEs, but also determines the overall
performance of the H-CRANs. Fronthaul with high bit rate and low
time latency enables larger-scale cooperative processing and
networking, which in turn makes the available scarce resources use
more efficiently. However, the practical fronthaul solutions have no
sufficient end-to-end performances to meet the desired QoS
requirements everywhere.

Considering that the non-ideal fronthaul may deteriorate overall SE
and EE performances of H-CRANs, where both the limited capacity and
time delay constraints are inevitable in practice, the solutions to
tackle these constraints should be given in the future. To overcome
the aforementioned problems related to the non-ideal fronthaul, new
cooperative processing and resource allocation strategies need to be
presented, and enhanced EE performance metric that considering the
constraints of fronthaul should be evolved. The optimal solution for
the resource allocation and admission control under the non-ideal
fronthaul constraint is an non-deterministic polynomial-time
(NP)-complete
problem\textcolor[rgb]{1.00,0.00,0.00}{\cite{bib:CRANcapacity}}.
Even in moderate dense RRHs deployment scenarios,
computational/memory costs may prevent H-CRANs from achieving an
optimal solution. Hence, the flexible iterative algorithms
characterized by limited complexity should be designed to improve
the overall SE and EE performances for H-CRANs in the future.

\subsection{Energy Harvesting in H-CRANs}

To save operational costs and decrease the energy consumptions, 5G
systems are anticipated with renewable energy supply, such as solar
and wind sources. Nevertheless, unlike the conventional energy from
grid, the renewable energy (e.g., harvested through solar panel or
wind turbine) is intermittent in nature and can have different
availabilities over time and space. Traditional methods like energy
storage with the use of capacity-limited and expensive battery are
far from enough for 5G systems to manage the fluctuations. To
mitigate such uncertain renewable energy fluctuations and shortages,
the energy cooperation by sharing the excessive energy among
RRHs/ACEs with the help of Node C is a promising solution. However,
one key problem that remains unaddressed yet to implement this
cooperation amongst large-scale ultra dense RRHs/ACEs is how to
motivate the nodes with renewable energy to share its excessive
energy to the others whose energy is insufficient with some benefits
in return. Furthermore, the energy cooperation will occupy the radio
resources, which will sacrifice the spectrum efficiency. To match
peak-hour wireless traffics with limited spectrum, RRHs and ACEs can
increase the transmission power to improve the spectrum efficiency,
which indicates that the energy harvesting should be adaptive to the
dynamical spectrum requirement and changeable traffic volume.

Most existing research works for optimizing the SE and EE
performances are assumed that the transmission power is fixed and
stable, however, based on the energy harvesting which has different
availabilities over time and space, these works should be researched
further. Summarily, the SE and EE performance optimizations in the
ultra dense RRHs/ACEs with energy harvesting in H-CRANs is a
promising and challenging work, where the transmission power of
nodes should be adaptive to the packet traffic, radio channel
fading, user's QoS, and offered energy.

\section{Conclusion}

In this article, we have provided a summary of recent advances in
the application scenario, system architecture, and key techniques
for achieving high throughput and low energy consumption in
heterogeneous cloud radio access networks (H-CRANs). To be
compatible with the developments of SDN and NFV in the fifth
generation wireless communication systems (5G), H-CRANs combine
advantages of both HetNets and C-RANs to act as the access network.
The key technologies, including advanced spatial signal processing,
cooperative radio resource management, network function
virtualization, and self-organizing network, are surveyed.
Meanwhile, potential challenges and open issues are discussed in
this article as well, including fronthaul constraints and their
corresponding performance optimization, and energy harvesting. The
presented key technologies and potential solutions to H-CRANs in
this article will be regarded as breakthroughs of the advanced next
generation wireless communication systems.

\section{Acknowledgment}

This work was supported in part by the National Natural Science
Foundation of China (Grant No. 61222103, No.61361166005), the
National High Technology Research and Development Program of China
(Grant No. 2014AA01A701), the National Basic Research Program of
China (973 Program) (Grant No. 2013CB336600), the State Major
Science and Technology Special Projects (Grant No. 2013ZX03001001),
the Beijing Natural Science Foundation (Grant No. 4131003).

\begin{IEEEbiography}{Mugen Peng}
(M'05--SM'11) received the PhD degree in Communication and
Information System from the Beijing University of Posts \&
Telecommunications (BUPT), China in 2005. Now he is a full professor
with the school of information and communication engineering in
BUPT. His main research areas include cooperative communication,
heterogeneous network, and cloud communication. He has
authored/coauthored over 40 refereed IEEE journal papers and over
200 conference proceeding papers. He received the 2014 IEEE ComSoc
AP Outstanding Young Researcher Award, and the Best Paper Award in
GameNets 2014, CIT 2014, ICCTA 2011, IC-BNMT 2010, and IET CCWMC
2009.
\end{IEEEbiography}

\begin{IEEEbiography}{Yong Li}(M'13) received the PhD degree in signal and information
processing from Beijing University of Posts and Telecommunications
(BUPT), Beijing, China, in 2009. He is currently an Associate
Professor with the School of Information and Communication
Engineering, BUPT. His current research interests include
cooperative communications and next-generation wireless networks.
\end{IEEEbiography}

\begin{IEEEbiography}{Zhongyuan Zhao} is currently a lecturer with the Key Laboratory of
Universal Wireless Communication (Ministry of Education) at Beijing University of Posts \& Telecommunications (BUPT), China.
He received his Ph.D. degree in communication and information systems and B.S. degree in applied mathematics from BUPT in 2009 and 2014, respectively.
His research interests include network coding, MIMO, relay transmissions, and large-scale cooperation in future communication networks.
\end{IEEEbiography}

\begin{IEEEbiography} {Chonggang Wang}
(SM'09) received his Ph.D. degree from BUPT in 2002. He is a member
technical staff with InterDigital Communications focusing on
Internet of Things (IoT) R \& D activities, including technology
development and standardization. His current research interests
include IoT, mobile communication and computing, and big data
management and analytics. He is the founding Editor-in-Chief of
\emph{IEEE Internet of Things Journal} and on the editorial board of
several journals, including IEEE Access.
\end{IEEEbiography}

\end{document}